# Six Interventions for the Responsible and Ethical Implementation of Medical AI Agents

Tom Bisson[1], Henriette Voelker[2,3], Sanddhya Jayabalan[4], A John Iafrate[1], Jakob N Kather[4,5,6,7], and Jochen K Lennerz[8]

[1]Department of Pathology, Massachusetts General Hospital, Harvard Medical School, Boston, MA, USA; [2]Charité – Universitätsmedizin Berlin, corporate member of Freie Universität Berlin, Humboldt Universität zu Berlin, Institute of Medical Informatics; [3]Charité – Universitätsmedizin Berlin, corporate member of Freie Universität Berlin, Humboldt Universität zu Berlin, Institute for the History of Medicine and Ethics in Medicine; [4]Else Kroener Fresenius Center for Digital Health, Faculty of Medicine, TUD Dresden University of Technology, Dresden, Germany; [5]Department of Medicine I, Faculty of Medicine, TUD Dresden University of Technology, Dresden, Germany; [6]Medical Oncology, National Center for Tumor Diseases (NCT), University Hospital Heidelberg, Heidelberg, Germany; [7]Pathology & Data Analytics, Leeds Institute of Medical Research at St James's, University of Leeds, Leeds, United Kingdom; [8]Natera, Inc., Austin, TX, USA

**Word count:** 1458 words

**References:** 5

**Figures:**    1, color

**Conflicts of Interest**: SJ declares part-time employment funded by AstraZeneca as a researcher at Heidelberg University; AJI receives royalties from ArcherDx/IDT and holds equity in Monimoi Therapeutics; JNK declares ongoing consulting services for AstraZeneca and Bioptimus. Furthermore, he holds shares in StratifAI, Synagen, and Spira Labs, has received an institutional research grant from GSK and AstraZeneca, as well as honoraria from AstraZeneca, Bayer, Daiichi Sankyo, Eisai, Janssen, Merck, MSD, BMS, Roche, Pfizer, and Fresenius; JKL is employed by Natera, Inc., Austin, TX, USA; TB and HV have no conflicting interests to declare.

**Corresponding Author:**
Joe Lennerz MD PhD
Natera, Inc., 13011A McCallen Pass
Austin, TX 78753, USA
E-Mail: JLennerz@natera.com

**ABSTRACT**

**Large language model (LLM)–based AI agents are increasingly capable of complex clinical reasoning and may soon participate in medical decision-making with limited or no real-time human oversight. This shift raises fundamental questions about how the core principles of medical ethics (i.e., beneficence, nonmaleficence, autonomy, and justice) can be upheld when the clinical responsibility extends to autonomous systems. Here we propose an *ethics-by-design* framework for medical AI agents comprising six practical interventions: auditable ethical reasoning modules, explicit human override conditions, structured patient preference profiles, AI-specific ethics oversight tools, global benchmarking repositories for ethical scenarios, and regulatory sandboxes for real-world evaluation. Together, these mechanisms aim to operationalize ethical governance for emerging clinical AI agents.**

https://github.com/BissonTom/Ethical-Governance-of-Medical-AI-Agents

**The Ethical Challenge of Clinical AI Agents**

AI agents based on *large language models* (LLMs; e.g., ChatGPT 5.2, Gemini 3.1, Claude 3.5) continue to evolve rapidly in healthcare. We are now confronted with a pressing question: can these systems, now capable of advanced reasoning with limited or no real-time human oversight, be held to the same ethical standards as human practitioners? Here, we focus on AI agents as autonomous systems capable of independent clinical decision-making [1] based on complex input data, often with limited or no real-time human oversight (=human out of the loop, HOOTL).

To manage this responsibility shift we argue for a clear framework for evaluating whether autonomous AI agents can uphold the four foundational principles of modern medical ethics: beneficence, nonmaleficence, autonomy, and justice [2]. LLMs can generate highly coherent language and have demonstrated complex reasoning across clinical and non-clinical domains. But performance must not be mistaken for competence [3]. These systems do not demonstrate self-awareness and have no intrinsic grasp of human goals, emotions, or ethical nuance. Without clear boundaries and governance structures, the growing role of AI in care delivery risks undermining the values defining medical practice.

Consider a fictional scenario: an FDA-authorized AI agent is tasked with managing a 67-year-old patient with metastatic pancreatic cancer. The patient has exhausted both first- and second-line therapies and expresses a clear wish for aggressive treatment, regardless of prognosis. After analysis of the information contained in the medical record, the AI concludes that third-line treatment offers minimal survival benefit with considerable toxicity, and due to the patient's declining performance status, recommends palliative care. Despite being presented with the patient's stated preferences, the AI deprioritized autonomy in favour of risk minimization. Now consider an alternate outcome. The AI administers third-line therapy as requested, resulting in severe toxicities and rapid clinical decline, ultimately leading to the patient's death. While this course respects autonomy, it fails to meet standards of beneficence and nonmaleficence. This hypothetical scenario highlights the inherent tension among ethical principles and underscores the challenge of encoding them into algorithmic decision-making processes.

The principle of justice further complicates this landscape, as it demands fair access to care while balancing individual needs with broader population-level resource allocation. It requires that individual care be balanced with equitable access across populations. The COVID-19 pandemic provided a stark illustration: limited ventilator availability forced clinicians to triage resources, sometimes withdrawing life-sustaining interventions from one patient to increase the likelihood of survival for another. These decisions require a delicate balance between maximizing individual outcomes and preserving system-wide function. AI agents may eventually be tasked with making similar judgments. Their reasoning could prioritize interventions with broader utility or lower cost, inadvertently shifting healthcare from a patient-cantered model to one driven by cost containment or operational efficiency. The tension between collective and individual ethics is culturally embedded – some societies emphasize communal welfare; others prioritize individual autonomy. Both approaches can be ethically justified, yet they represent fundamentally different moral frameworks.

LLMs can be tuned or otherwise constrained to reflect these differing cultural, religious, and societal perspectives — but alignment does not equate to comprehension [4]. We believe that for AI agents to serve ethically in healthcare, they should be guided by more than computational optimization. We recommend constraining and evaluating them within an ethics-by-design framework–one that proactively embeds our core ethical values and institutional responsibilities into the development, deployment, and oversight of AI systems.

Here, we therefore propose a principled ethical framework (Figure 1) specific to AI agents in healthcare–integrating existing bioethical standards while addressing the unique challenges posed by autonomous reasoning systems.

**Six key interventions**

**1. – Ethical Reasoning Module with Auditable Interactions**
Every AI agent should consult an Ethical Reasoning Module (ERM), designed to review the medical AI agent's decisions from an ethical viewpoint and engage in mutual discussion to minimize ethical tension. The ERM may be integrated within the AI agent, operate as a standalone software module, or be engaged via third-party vendor services. It must produce a structured, auditable record of interactions–including inputs, outputs, reasoning traces, and intermediate agent exchanges–that document the ethical trade-offs considered. The ERM should be subject to regulatory approval by appropriate governing bodies.

**2. – Explicit Human Override Conditions Based on Clinical Context**
AI agent decisions should be revisable, questionable, and rejectable by a human-in-the-loop (e.g., practitioner, patient, proxy), whose identity must be recorded to address legal accountability and clinical responsibility. Explicit human override conditions should be defined and human intervention triggered, when predefined clinical thresholds (e.g., high-risk interventions, irreversible outcomes) are exceeded.

**3. – Structured Patient Preference Profiles with Dynamic Updating**
AI agents should be designed to incorporate a patient's individual background information, ethical frameworks, beliefs, and wishes. *Structured Patient Preference Profiles* (SPPPs), composed of advance directives, values surveys, and optional ethical stance questionnaires may serve as an initial source for the AI agent but real-time feedback facilitating dynamic adaptation to acute circumstances must not become dispensable (see https://github.com/BissonTom/Ethical-Governance-of-Medical-AI-Agents).

**4. – AI-Specific Tools and Training for Institutional Ethics Committees**
Institutional ethics committees should adapt to assess AI Agents' decisions with the same rigor they use to review human practitioners' actions. Documented ethical trade-offs within AI agents may be excessive, necessitating maintained domain-specific review tools, structured evaluation templates and training to support meaningful review and reduce administrative burden. To remain effective, oversight tools must be regularly updated and streamlined to avoid bureaucratic bottlenecks while maintaining rigorous, usable governance as AI integrates into clinical workflows.

**5. – Global Open-Access Repository for Ethical Scenario Benchmarking**
A centralized, open-access repository of diverse real-world and fictional ethical scenarios should be curated and maintained by a neutral, representative international consortium. Contributions should be accepted by different stakeholders across the globe to facilitate democratized participation. Benchmarking tools for AI agents should be developed to provide a structured, measurable and longitudinal evaluation of ethical reasoning capabilities across AI systems in both development and after deployment as quality control and proficiency testing, enforced by national regulatory bodies. The benchmark should include a structured update mechanism to reflect ongoing advances in AI capabilities.

### 6. – Regulatory Sandboxes for Real-World Ethical Testing

Medical AI agents should be evaluated in regulatory sandboxes: time-limited, real-world deployments authorized under regulatory supervision. These environments allow AI systems to operate in clinical settings while being closely monitored for ethical performance, patient impact, and safeguard alignment. Sandboxes provide a structured way to detect and correct ethical risks before full-scale rollout. They enable regulators, developers, and clinicians to collaborate in refining agent behaviour, documenting decision-making, and ensuring accountability. Shared guidance on sandbox criteria would support transparency, comparability, and responsible innovation across health systems.

## Operationalizing Ethical Reasoning

Most agentic frameworks primarily address the agentic AI function with few exceptions incorporating regulatory concerns and establishing concrete approaches to agent-level compliance [5]. The interventions presented here aim to directly integrate ethical perspectives into the operational logic of AI agents. We point out that the six interventions follow a dynamic design principle that allows for continuous updates as technologies or requirements evolve. We propose the term *ethics-by-design* to account for the need to adapt. In other words, agentic AI is confronted with an unprepared healthcare system. The barriers go beyond technical limitations. They also go beyond systematic review of the LLM output [6], and beyond clinical simulations [5]. Here we argue that agentic AI demands an adaptive ethical governance structure – a structure with benefits that extend to the dynamic reality of human decision-making. Healthcare will thereby be able to account for the complexities of provider-patient interactions and the right of self-determination reflected in individual ethical perspectives.

Recent developments underscore the urgent need to establish and implement ethical guiding principles for AI in healthcare. In early 2025, for example, the U.S. House of Representatives introduced the Healthy Technology Act, a bill that includes language recognizing artificial intelligence (AI) and machine learning systems as eligible practitioners with prescribing authority [7]. We are entering a world in which AI systems may be treated as de facto clinical decision-makers. Therefore, national supervisory authorities must adapt existing accountability mechanisms to ensure they are applicable and enforceable for AI systems. While numerous ethical and corresponding regulatory frameworks have been proposed, the next step is to establish legal mechanisms that both enforce and incentivize adherence to ethical principles by AI agents.

We call on regulatory and policy experts to integrate these six interventions into governance. We invite healthcare professionals and patient advocates to contribute real-world dilemmas for benchmarking and urge funders to support these efforts. And we hope to inspire computer scientists and the open-source community to develop the technical infrastructure needed to translate ethical intent into operational design for AI Agents in healthcare. We must align across policy, practice and technology to ensure AI agents act as ethical collaborators in care — not a silent replacement for human judgment.

**Figure**

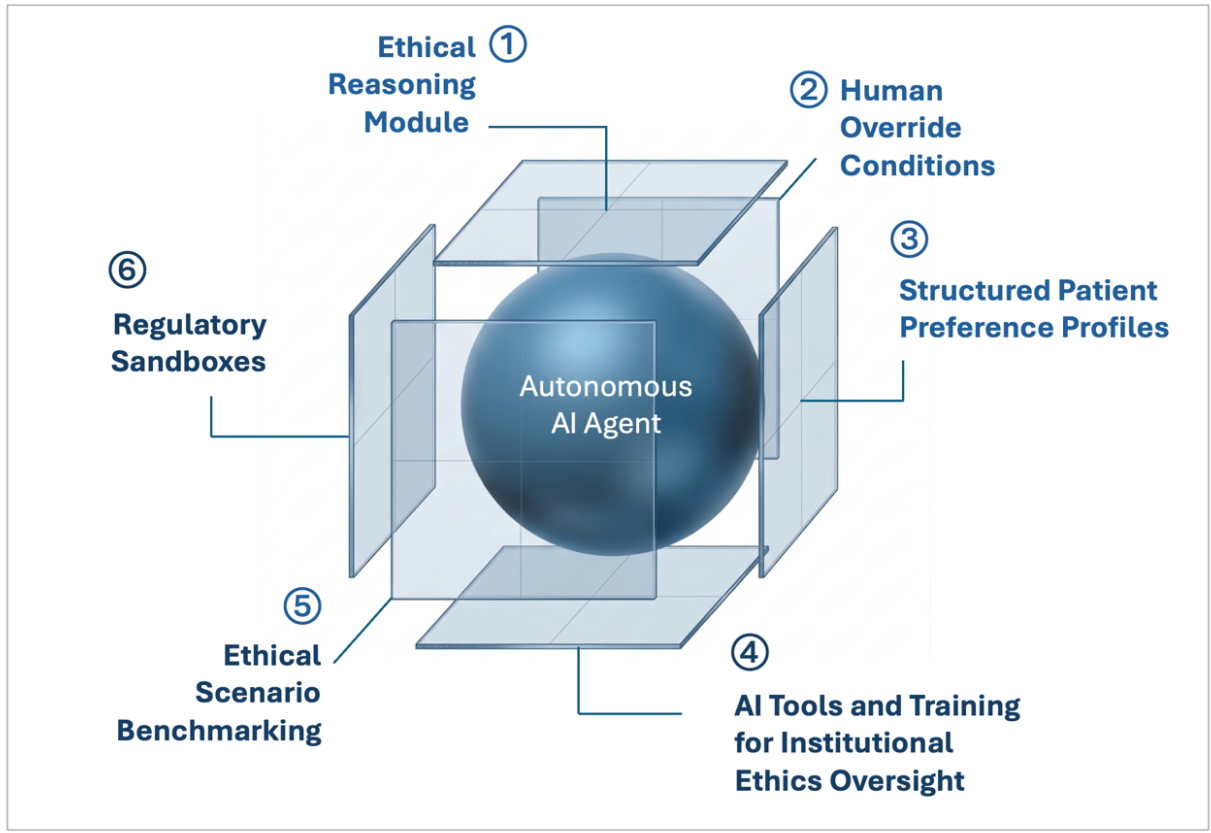

**Figure 1: Ethics-by-Design Framework for Medical AI Agents.**
The central sphere represents the autonomous AI agent capable of clinical decision-making. It is encapsulated by a six-plane boundary system representing the proposed ethical interventions: an Ethical Reasoning Module for auditable reflection; Human Override Conditions to ensure accountability; Structured Patient Preference Profiles for dynamic alignment with individual values; Institutional Ethics Oversight via updated review tools; Ethical Scenario Benchmarking using a global open-access repository; and Regulatory Sandboxes for supervised real-world testing. Together, these planes constrain the agent to ensure adherence to the foundational principles of beneficence, nonmaleficence, autonomy, and justice.